\documentclass[a4paper]{report}
\usepackage[utf8]{inputenc}
\usepackage[T1]{fontenc}
\usepackage{RJournal}
\usepackage{amsmath,amssymb,array}
\usepackage{booktabs}

\providecommand{\tightlist}{%
  \setlength{\itemsep}{0pt}\setlength{\parskip}{0pt}}

\newlength{\cslhangindent}
\newlength{\csllabelwidth}
\newlength{\cslentryspacingunit} 
  {}%
  {\par}
\newenvironment{CSLReferences}[2] 
 {
  \setlength{\parindent}{0pt}
  \ifodd #1
  \let\oldpar\par
  \def\par{\hangindent=\cslhangindent\oldpar}
  \fi
  \setlength{\parskip}{#2\cslentryspacingunit}
 }%
 {}
\usepackage{calc}

\usepackage{longtable}
\definecolor{link}{RGB}{39,109,195}
\begin{document}

\begin{article}

\title{CRAN Task Views: The Next Generation}
\author{by Achim Zeileis, Roger Bivand, Dirk Eddelbuettel, Kurt Hornik, and Nathalie Vialaneix}

\maketitle

\abstract{%
CRAN Task Views have been available on the Comprehensive R Archive Network since 2005. They provide guidance about which CRAN packages are relevant for tasks related to a certain topic, and can also facilitate automatic installation of all corresponding packages. Motivated by challenges from the growth of CRAN and the R community as a whole since 2005, all of the task views infrastructure and workflows were rethought and relaunched in 2021/22 in order to facilitate maintenance and to foster deeper interactions with the R community. The redesign encompasses the establishment of a group of CRAN Task View Editors, moving all task view sources to dedicated GitHub repositories, adopting well-documented workflows with a code of conduct, and leveraging R/Markdown files (rather than XML) for the content of the task views.
}

\hypertarget{motivation-and-background}{%
\section{Motivation and background}\label{motivation-and-background}}

The Comprehensive R Archive Network (CRAN, \url{https://CRAN.R-project.org/}), comprised
of about hundred mirror pages around the world along with a content-delivery network (CDN), is the canonical site
for downloading not only the base R system but also about twenty thousand extension packages
(actually, 19565 at the time of writing), contributed by users from
the R community.  CRAN is a very useful resource within the R
ecosystem, but its content can be hard to grasp and navigate. Especially users new to R
can struggle with getting started and finding the relevant R packages for the
tasks they want to accomplish.

To mitigate this problem, Zeileis (2005), in cooperation with the CRAN team, introduced
so-called \emph{CRAN Task Views} that provide guidance about which CRAN packages
are relevant for tasks related to a given topic. Task views offer a brief overview of
the included packages on a dedicated CRAN page (see
\url{https://CRAN.R-project.org/web/views/} for an overview) and also enable the
automatic installation of all packages from a task view using the
\CRANpkg{ctv} package (Zeileis and Hornik 2022). The views are intended to have a sharp focus so
that it is sufficiently clear which packages should be included (or excluded).
They are not meant to endorse the ``best'' packages for a given task but they can
distinguish a shorter list of \emph{core} packages that are most relevant/important from the
remaining \emph{regular} packages.

While CRAN Task Views alone were certainly not able to overcome the problems of
navigating CRAN, they proved to be useful enough to be continued over the subsequent
years. However, due to the growth of CRAN (see, e.g., Hornik, Ligges, and Zeileis 2022) and of the R community as a whole
since the introduction of task views in 2005, the thriving of the task views
was limited by several design decisions regarding their format and the corresponding
workflows:

\begin{itemize}
\tightlist
\item
  The format for authoring a new task view was based on XML (extensible markup
  language). This required that task view maintainers as well as anyone who wanted to
  contribute write XML and HTML (hypertext markup language) directly.
\item
  The format required that the packages included in a task view needed to be
  described in the information text and listed separately to be categorized
  into \emph{core} and \emph{regular} packages.
\item
  Task views were typically proposed by individual contributors (and only in some cases
  by teams).
\item
  The onboarding process for CRAN was mostly coordinated by Achim Zeileis alone.
\item
  All task views were maintained in a single subversion (SVN) repository on the
  R-Forge platform (Theußl and Zeileis 2009) to which all task view maintainers had access.
\item
  Contributions from the R community were mostly limited to e-mails to the respective
  maintainers (except where the maintainers set up other channels of communication
  themselves, e.g., through a separate GitHub repository).
\end{itemize}

This setup worked sufficiently well in 2005 when there were about 500~packages on CRAN and
initially 4~task views, listing on average around 20~packages each. (Within a year, the
number of task views increased to 12.)

But by now there are 43~task views, with median and mean numbers of CRAN
packages covered 106 and~121, respectively.
Overall, these task views cover 4360~CRAN packages,
which is about 22\% of all CRAN packages.

This change in scope and scale necessitated a change in infrastructure and workflows
that the new \emph{CRAN Task View Initiative}, described in detail in the next section,
aims to provide.

\hypertarget{the-cran-task-view-initiative}{%
\section{The CRAN Task View Initiative}\label{the-cran-task-view-initiative}}

Motivated by the challenges from the growth of CRAN and the R community as a whole,
almost all aspects of the CRAN Task Views were rethought and relaunched in 2021/22
in order to facilitate maintenance and foster more interactions with the R community.
The important changes are:

\begin{itemize}
\tightlist
\item
  The new \emph{CRAN Task View Initiative} is overseen by a group of \emph{CRAN Task View Editors}
  (rather than an individual)
  who review proposals of new task views, support the onboarding of the corresponding
  maintainers, and monitor the activity in existing task views. The corresponding
  official e-mail is \texttt{CRAN-task-views@R-project.org}.
\item
  All activities are hosted on GitHub (rather than R-Forge) in a dedicated organization
  (\url{https://github.com/cran-task-views/}) which provides interfaces and workflows that
  many R users are familiar with. In particular, it offers a wide range of possibilities
  for the community to engage with the task views, most notably through issues and
  pull requests.
\item
  Each task view is hosted in a separate repository within the \texttt{cran-task-views}
  organization (rather than all in one repository), giving the maintainers more freedom
  while preserving sufficient control for the editors. Also, separate projects
  provide better visibility for each task view and a clearer separation of responsibilities.
\item
  For new task view proposals, the principal maintainer of a task view is expected to assemble
  a team of 1-5 co-maintainers to share the workload and reflect different perspectives. The
  same was also strongly encouraged for older task views that had previously been maintained by a single
  person. Ideally the co-maintainers should be a diverse group in terms of gender,
  origin, scientific field, etc.
\item
  The file format for authoring task views is now based on R/Markdown (rather than XML
  and HTML directly). The files can be processed and rendered into HTML output for CRAN
  with dedicated functions from the \pkg{ctv} package, leveraging the popular packages
  \CRANpkg{knitr} (Xie 2015) and \CRANpkg{rmarkdown} (Xie, Allaire, and Grolemund 2018).
\item
  As the task view files are now dynamic R/Markdown documents, it was easy to avoid
  certain redundancies from the old XML-based format, e.g., the package list with \emph{core}
  and \emph{regular} packages is compiled automatically from the information text and does
  not have to be specified separately.
\item
  All contributions must now explicitly adhere to the
  \href{https://github.com/cran-task-views/ctv/blob/main/CodeOfConduct.md}{code of conduct}
  of the initiative, adapted from the the \href{https://www.contributor-covenant.org/}{Contributor Covenant}
  code of conduct.
\end{itemize}

More details on the CRAN Task View Initiative are available in the GitHub repository
\url{https://github.com/cran-task-views/ctv}. This provides an overview of the activities,
detailed documentation, and the possibility to raise issues that concern the initiative
as a whole (rather than individual task views).

The current CRAN Task View Editors are
\href{https://github.com/rsbivand}{Roger Bivand},
\href{https://github.com/eddelbuettel}{Dirk Eddelbuettel},
\href{https://github.com/tuxette}{Nathalie Vialaneix}, and
\href{https://github.com/zeileis}{Achim Zeileis}.
Former contributors include: Henrik Bengtsson, Rocío Joo, David Meyer, and Heather Turner.
The primary contributor from the \href{https://CRAN.R-project.org/CRAN_team.htm}{CRAN team} is
\href{https://github.com/kurthornik}{Kurt Hornik}.

\hypertarget{using-task-views}{%
\section{Using task views}\label{using-task-views}}

First and foremost, the web page of each task view can be read by interested users
(see Figure~1 for an example).
The clear structure and focus of the task views should help the readers to quickly
gain a first overview of a topic and to search for specific relevant packages
more efficiently.

\begin{figure}[t!]
\includegraphics[width=1\linewidth]{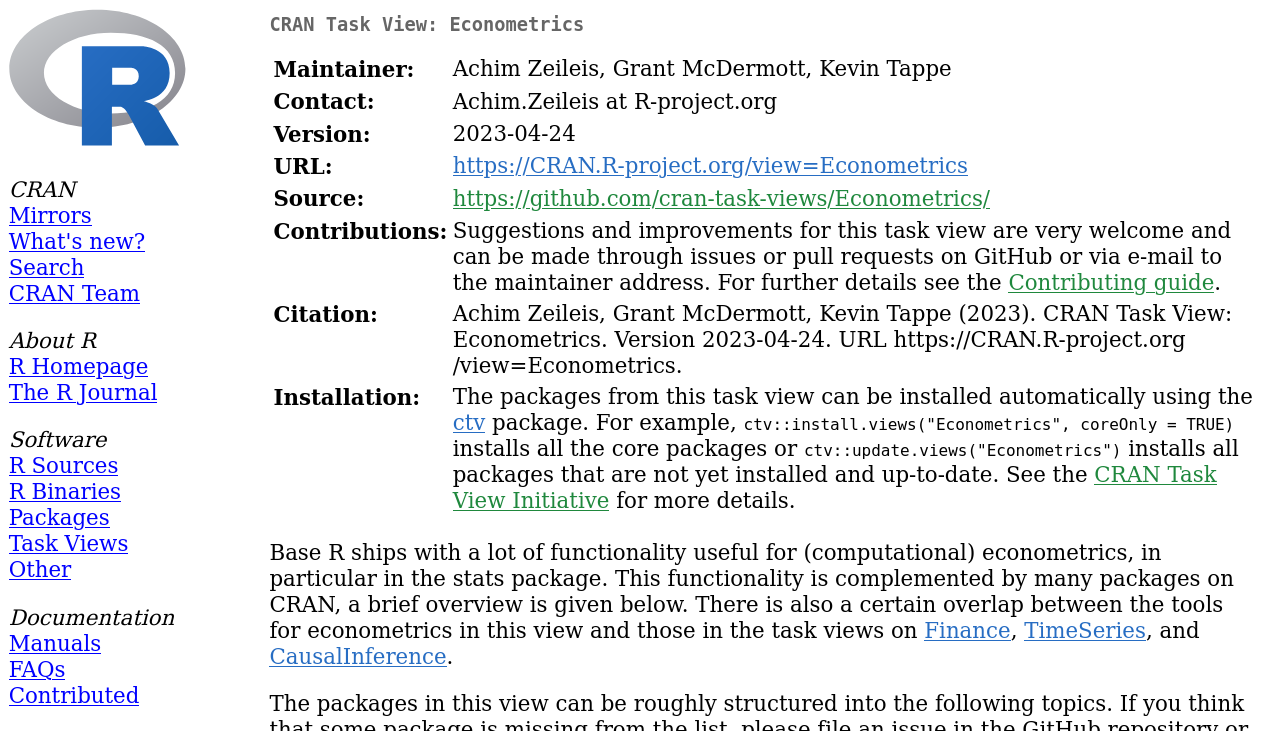} \caption{Screenshot of the header and introduction of the "Econometrics" task view.}\label{fig:econometrics}
\end{figure}

Another benefit of task views that has probably been underappreciated for a long time is
that they allow the easy installation of the associated packages (either all of them or
just the core packages). The headers of all task view web pages now promote this
possibility explicitly. To make use of this, the \href{https://CRAN.R-project.org/package=ctv}{ctv}
package needs to be installed, e.g., via

\begin{verbatim}
install.packages("ctv")
\end{verbatim}

The package provides the two functions \texttt{install.views()} and \texttt{update.views()},
where the latter only installs those packages which are not installed and up-to-date.
For example, in order to install the full ``Econometrics'' task view either one of the
following two calls can be used:

\begin{verbatim}
ctv::install.views("Econometrics")
ctv::update.views("Econometrics")
\end{verbatim}

Moreover, two functions are provided for querying the information from task views
from within~R. First, information on a single task view can be obtained with the
\texttt{ctv()} function.

\begin{verbatim}
ctv::ctv("Econometrics")
\end{verbatim}

\begin{verbatim}
#> CRAN Task View
#> --------------
#> Name:       Econometrics
#> Topic:      Econometrics
#> Maintainer: Achim Zeileis, Grant McDermott, Kevin Tappe
#> Contact:    Achim.Zeileis@R-project.org
#> Version:    2023-04-24
#> Repository: https://CRAN.R-project.org
#> Source:     https://github.com/cran-task-views/Econometrics/
#> Packages:   AER*, alpaca, aod, apollo, apt, bacondecomp, bayesm, betareg, bife,
#>             bimets, BMA, BMS, boot, bootstrap, brglm, CADFtest, car*, CDNmoney,
...
\end{verbatim}

To list all task views available from CRAN, the function \texttt{available.views()} can
be used:

\begin{verbatim}
ctv::available.views()
\end{verbatim}

\begin{verbatim}
#> CRAN Task Views
#> ---------------
#> Name: Agriculture
#> Topic: Agricultural Science
#> Maintainer: Julia Piaskowski, Adam Sparks, Janet Williams
#> Repository: https://CRAN.R-project.org
#> ---------------
#> Name: Bayesian
#> Topic: Bayesian Inference
#> Maintainer: Jong Hee Park, Michela Cameletti, Xun Pang, Kevin M. Quinn
#> Repository: https://CRAN.R-project.org
#> ---------------
...
\end{verbatim}

The objects returned by \texttt{ctv()} and \texttt{available.views()} include additional
information which is not shown by the \texttt{print()} method. This may be useful
for more specific computations based on the task views. For example, a
\texttt{citation} object (Hornik, Murdoch, and Zeileis 2012) can be obtained from the list returned by \texttt{ctv()}.

\begin{verbatim}
ctv("Econometrics")$citation
\end{verbatim}

\begin{verbatim}
#> To cite the Econometrics task view in publications use:
#> 
#>   Zeileis A, McDermott G, Tappe K (2023). _CRAN Task View:
#>   Econometrics_. Version 2023-04-24,
#>   <https://CRAN.R-project.org/view=Econometrics>.
#> 
#> A BibTeX entry for LaTeX users is
#> 
#>   @Manual{,
#>     author = {Achim Zeileis and Grant McDermott and Kevin Tappe},
#>     title = {CRAN Task View: Econometrics},
#>     year = {2023},
#>     note = {Version 2023-04-24},
#>     url = {https://CRAN.R-project.org/view=Econometrics},
#>   }
\end{verbatim}

\hypertarget{contributing}{%
\section{Contributing}\label{contributing}}

Users from the R community can contribute in two ways to the CRAN
Task View Initiative: They can either provide suggestions for an \emph{existing}
task view, or they can \emph{propose a new one} on a topic that is not yet covered.
In either case, all contributions must be made under the
\href{https://github.com/cran-task-views/ctv/blob/main/CodeOfConduct.md}{code of conduct}.

Contributions to existing task views are welcome, encouraged, and in fact crucial for
keeping the task views up to date. Typical contributions would be improvements
in existing content (e.g., adding details, clarifications, or corrections) or
suggestions of additional content (e.g., packages or links). To facilitate such
contributions, each task view includes the e-mail address of the principal
maintainer as well as a link to the associated GitHub repository. Thus, contributors
can choose the most convenient alternative among the following ones:

\begin{itemize}
\tightlist
\item
  \emph{Send an e-mail} to the principal maintainer.
\item
  \emph{Raise an issue} in the GitHub repository.
\item
  \emph{Provide a pull request} in the GitHub repository.
\end{itemize}

To avoid having a pull request become more involved or disruptive, it is
frequently suggested to first discuss proposed changes by raising an issue,
and to make sure that the modified task view
file still works correctly (see Section~\protect\hyperlink{sec:format}{6}).

For proposing new task views, the CRAN Task View Initiative provides a standardized
workflow including a review and onboarding process. All steps are detailed at
\url{https://github.com/cran-task-views/ctv/blob/main/Proposal.md}. In the following,
we just outline the essential aspects.

First, it is important that the prospective maintainers consider the time and work
that is required to put together a proposal, to refine it during the review and
onboarding process, and most importantly to actively maintain the task view in the
future.

Second, a fundamental and very crucial step is to formulate the topic of the new
task view so that it has a clear scope that is neither too narrow nor too wide.
The goal is not to cover ``every package'' remotely related to the topic but rather
the set of packages that clearly fall within the scope. The coverage should be similar to
what an (introductory) text book on the topic would cover. Non-CRAN packages may also
be included but the focus of CRAN task views should be packages on CRAN (as the name conveys).
Finally, task views should \emph{not} rate the packages or endorse certain ``best'' packages
but rather give an overview of what is available. A bit of emphasis to the more
important packages can be given in two ways: (a) The most important packages can be
flagged as \emph{core} packages. (b) In the information text the more important packages
can be listed first in the respective sections.

Third, based on the information formulated as outlined above, the proposal can be
made in the GitHub issues of the \texttt{ctv} repository. This initiates a first review
that is carried out in the comments of the issue tracker which are open to all
members of the community. In all cases, the CRAN Task View Editors are explicitly
invited to comment, as may be maintainers of related task views or of relevant core
packages etc. Provided that there is sufficient endorsement from the CRAN Task View
Editors, typically after revisions and refinements from the prospective maintainers,
a proposal is accepted, initiating an onboarding process that leads to the publication
of the task view on CRAN.

Finally, the maintainers of the task view are responsible for keeping it up to date
by checking CRAN regularly. Contributions from the community, as described at the
beginning of this section, are eminently useful for this and hence explicitly
encouraged. Moreover, some R packages like \pkg{CTVsuggest} (Dijk 2023) can support
the maintainers in discovering new relevant packages on CRAN.

\hypertarget{handling-package-archivals}{%
\section{Handling package archivals}\label{handling-package-archivals}}

The CRAN packages listed in task views should ideally be maintained actively,
so that improved versions are released by the corresponding maintainers in case
the daily CRAN checks discover any issues.
However, it is not straightforward to test for active maintenance fully automatically
and even actively maintained packages may be temporarily archived on
CRAN. Hence, the following strategy is adopted within the CRAN Task View
Initiative: When a CRAN package from a task view is archived, it is still listed in the
task view like before. It is only flagged as archived in the text and not
installed automatically anymore by \texttt{install.views()} and \texttt{update.views()}.

This strategy gives both the package maintainers and the task view maintainers
some time to resolve the situation. Specifically, the task view maintainers can
decide whether to

\begin{itemize}
\tightlist
\item
  \emph{exclude} the package from the task view immediately, e.g., if it was
  archived for policy violations, at the request of the maintainer, or did not
  have any updates for many years and is not associated with a public repository;
\item
  \emph{wait some more} for an improved version, e.g., when they see that the
  package maintainers already started addressing the problem; or
\item
  \emph{reach out to the package maintainers} to check if they intend to release a
  corrected version or even to help with releasing an improved
  version.
\end{itemize}

To help discovering archived packages and initiating one of the actions above,
the CRAN team regularly checks whether any task view contains packages that have
been archived on CRAN for 60~days or more. If so, they create an issue in the
corresponding task view repository.

After the period of grace (100 days) ends, the situation should be resolved by
the task view maintainers, typically by excluding packages that are still archived from the
task view. For sufficiently relevant packages, it may be sensible to replace the
package listing by a link, e.g., to a GitHub repository for the package.

\hypertarget{sec:format}{%
\section{R/Markdown format}\label{sec:format}}

The file format for CRAN task views leverages the R/Markdown format (Xie, Allaire, and Grolemund 2018)
so that standard Markdown can be used for formatting and structuring the
text and a handful of special R functions are provided to link to CRAN
packages, other task views, GitHub projects, etc.
The format is mostly self-explanatory and is illustrated below using an
excerpt from the \texttt{Econometrics} task view:

\begin{verbatim}
---
name: Econometrics
topic: Econometrics
maintainer: Achim Zeileis, Grant McDermott, Kevin Tappe
email: Achim.Zeileis@R-project.org
version: 2022-09-13
source: https://github.com/cran-task-views/Econometrics/
---

Base R ships with a lot of functionality useful for (computational) econometrics,
in particular in the stats package. This functionality is complemented by many
packages on CRAN, a brief overview is given below. There is also a certain
overlap between the tools for econometrics in this view and those in the task
views on `r view("Finance")`, `r view("TimeSeries")`, and
`r view("CausalInference")`.

Further information can be formatted with standard Markdown syntax, e.g., for
_emphasizing text_ or showing something really important in **bold face**.
R/Markdown syntax with special functions can be used to link to a standard
package like `r pkg("mlogit")` or an important "core" package like
`r pkg("AER", priority = "core")`.

### Links
- [The Title of a Relevant Homepage](http://path/to/homepage/)
\end{verbatim}

The document structure consists of three main blocks: (a) Some metainformation
is given in the YAML header at the beginning (separated by lines with \texttt{-\/-\/-}),
followed by (b) the information in the main text, and (c) a concluding special
section called \texttt{\#\#\#\ Links}. Details are explained in the official documentation:
\url{https://github.com/cran-task-views/ctv/blob/main/Documentation.md}.

The information in the main text should be a short description of the packages,
explaining which packages are useful for which tasks. Additionally, short R code
chunks with special functions are used for linking to CRAN resources:
\texttt{pkg()} for regular packages, \texttt{pkg(...,\ priority\ =\ "core")} for important \emph{core}
packages, and \texttt{view()} for related task views. Moreover, code projects in other
repositories can be linked by dedicated functions, e.g., \texttt{bioc()} or
\texttt{github()} for packages on Bioconductor or GitHub, respectively.

In order to check whether a task view file has been formatted properly it can be read into
R and rendered to HTML (see also Figure~1) which can be opened and inspected in a browser. Additionally,
the function \texttt{check\_ctv\_packages()} can be used to check whether some of the listed
packages are actually not available on CRAN or not currently maintained
(archived). The functions are illustrated below, assuming that the \texttt{Econometrics.md}
file is in the local working directory:

\begin{verbatim}
ctv::ctv2html("Econometrics.md", cran = TRUE)
browseURL("Econometrics.html")
ctv::check_ctv_packages("Econometrics.md")
\end{verbatim}

Note that the extension \texttt{.md} (rather than \texttt{.Rmd}) has been adopted for the files
so that GitHub renders a Markdown preview on the fly. Finally, in case there are
still task view files that employ the legacy XML format, there is the (unexported)
function \texttt{ctv:::ctv\_xml\_to\_rmd()} that facilitates the transition to the new
R/Markdown format.

\hypertarget{available-task-views}{%
\section{Available task views}\label{available-task-views}}

Currently, there are 43~task views on CRAN with the following names
and maintainers:

\begin{itemize}
\tightlist
\item
  \href{https://CRAN.R-project.org/view=Agriculture}{\emph{Agriculture}} (Piaskowski, Sparks, Williams).
\item
  \href{https://CRAN.R-project.org/view=Bayesian}{\emph{Bayesian}} (Park, Cameletti, Pang, Quinn).
\item
  \href{https://CRAN.R-project.org/view=CausalInference}{\emph{CausalInference}} (Mayer, Zhao, Greifer, Huntington-Klein, Josse).
\item
  \href{https://CRAN.R-project.org/view=ChemPhys}{\emph{ChemPhys}} (Mullen).
\item
  \href{https://CRAN.R-project.org/view=ClinicalTrials}{\emph{ClinicalTrials}} (Zhang, Zhang, Zhang).
\item
  \href{https://CRAN.R-project.org/view=Cluster}{\emph{Cluster}} (Leisch, Gruen).
\item
  \href{https://CRAN.R-project.org/view=Databases}{\emph{Databases}} (Tang, Balamuta).
\item
  \href{https://CRAN.R-project.org/view=DifferentialEquations}{\emph{DifferentialEquations}} (Petzoldt, Soetaert).
\item
  \href{https://CRAN.R-project.org/view=Distributions}{\emph{Distributions}} (Dutang, Kiener, Swihart).
\item
  \href{https://CRAN.R-project.org/view=Econometrics}{\emph{Econometrics}} (Zeileis, McDermott, Tappe).
\item
  \href{https://CRAN.R-project.org/view=Environmetrics}{\emph{Environmetrics}} (Simpson).
\item
  \href{https://CRAN.R-project.org/view=Epidemiology}{\emph{Epidemiology}} (Jombart, Rolland, Gruson).
\item
  \href{https://CRAN.R-project.org/view=ExperimentalDesign}{\emph{ExperimentalDesign}} (Groemping, Morgan-Wall).
\item
  \href{https://CRAN.R-project.org/view=ExtremeValue}{\emph{ExtremeValue}} (Dutang).
\item
  \href{https://CRAN.R-project.org/view=Finance}{\emph{Finance}} (Eddelbuettel).
\item
  \href{https://CRAN.R-project.org/view=FunctionalData}{\emph{FunctionalData}} (Scheipl, Arnone, Hooker, Tucker, Wrobel).
\item
  \href{https://CRAN.R-project.org/view=GraphicalModels}{\emph{GraphicalModels}} (Hojsgaard).
\item
  \href{https://CRAN.R-project.org/view=HighPerformanceComputing}{\emph{HighPerformanceComputing}} (Eddelbuettel).
\item
  \href{https://CRAN.R-project.org/view=Hydrology}{\emph{Hydrology}} (Albers, Prosdocimi).
\item
  \href{https://CRAN.R-project.org/view=MachineLearning}{\emph{MachineLearning}} (Hothorn).
\item
  \href{https://CRAN.R-project.org/view=MedicalImaging}{\emph{MedicalImaging}} (Whitcher, Clayden, Muschelli).
\item
  \href{https://CRAN.R-project.org/view=MetaAnalysis}{\emph{MetaAnalysis}} (Dewey, Viechtbauer).
\item
  \href{https://CRAN.R-project.org/view=MissingData}{\emph{MissingData}} (Josse, Mayer, Tierney, Vialaneix).
\item
  \href{https://CRAN.R-project.org/view=MixedModels}{\emph{MixedModels}} (Bolker, Piaskowski, Tanaka, Alday, Viechtbauer).
\item
  \href{https://CRAN.R-project.org/view=ModelDeployment}{\emph{ModelDeployment}} (Tang, Balamuta).
\item
  \href{https://CRAN.R-project.org/view=NaturalLanguageProcessing}{\emph{NaturalLanguageProcessing}} (Wild).
\item
  \href{https://CRAN.R-project.org/view=NumericalMathematics}{\emph{NumericalMathematics}} (Borchers, Hankin, Sokol).
\item
  \href{https://CRAN.R-project.org/view=OfficialStatistics}{\emph{OfficialStatistics}} (Templ, Kowarik, Schoch).
\item
  \href{https://CRAN.R-project.org/view=Omics}{\emph{Omics}} (Aubert, Hocking, Vialaneix).
\item
  \href{https://CRAN.R-project.org/view=Optimization}{\emph{Optimization}} (Schwendinger, Borchers).
\item
  \href{https://CRAN.R-project.org/view=Pharmacokinetics}{\emph{Pharmacokinetics}} (Denney).
\item
  \href{https://CRAN.R-project.org/view=Phylogenetics}{\emph{Phylogenetics}} (Gearty, O'Meara, Berv, Ballen, Ferreira, Lapp, Schmitz, Smith, Upham, Nations).
\item
  \href{https://CRAN.R-project.org/view=Psychometrics}{\emph{Psychometrics}} (Mair, Rosseel, Gruber).
\item
  \href{https://CRAN.R-project.org/view=ReproducibleResearch}{\emph{ReproducibleResearch}} (Blischak, Hill, Marwick, Sjoberg, Landau).
\item
  \href{https://CRAN.R-project.org/view=Robust}{\emph{Robust}} (Maechler).
\item
  \href{https://CRAN.R-project.org/view=Spatial}{\emph{Spatial}} (Bivand, Nowosad).
\item
  \href{https://CRAN.R-project.org/view=SpatioTemporal}{\emph{SpatioTemporal}} (Pebesma, Bivand).
\item
  \href{https://CRAN.R-project.org/view=SportsAnalytics}{\emph{SportsAnalytics}} (Baumer, Nguyen, Matthews).
\item
  \href{https://CRAN.R-project.org/view=Survival}{\emph{Survival}} (Allignol, Latouche).
\item
  \href{https://CRAN.R-project.org/view=TeachingStatistics}{\emph{TeachingStatistics}} (Northrop).
\item
  \href{https://CRAN.R-project.org/view=TimeSeries}{\emph{TimeSeries}} (Hyndman, Killick).
\item
  \href{https://CRAN.R-project.org/view=Tracking}{\emph{Tracking}} (Joo, Basille).
\item
  \href{https://CRAN.R-project.org/view=WebTechnologies}{\emph{WebTechnologies}} (Sepulveda, Beasley).
\end{itemize}

\hypertarget{outlook}{%
\section{Outlook}\label{outlook}}

The new CRAN Task View Initiative has redesigned and relaunched the infrastructure
and workflows for CRAN Task Views so that they can thrive in the years to come. In particular,
many tools are used that are well-established in the R community such as \pkg{knitr}/\pkg{rmarkdown}
or collaborations through GitHub projects. Moreover, various steps have been taken in
order to assure that all task views are actively maintained and foster contributions
from the community, either in terms of additions/improvements for existing task views,
or in the form of new proposals. Since announcing the new initiative in Spring 2022,
many task views were already improved, e.g., by adding new content, extending the maintainer
teams, or incorporating feedback from the community. Additionally, there were already seven
successful new task view proposals (Agriculture, CausalInference, Epidemiology, MixedModels, Omics, Phylogenetics, SportsAnalytics).

While these steps already accomplished important improvements in the initiative,
further challenges remain for the future. Apart from improving the breadth and depth of
the task views, the most important aim is probably to better connect with those R users who
would profit from the information provided in the task views. We feel that this was easier
in the mid-2000s for two reasons:

\begin{enumerate}
\def\labelenumi{\arabic{enumi}.}
\item
  CRAN (and hence its task views) were actively browsed/searched by many R users, whereas
  today many more users will expect that useful content is presented to them, e.g., via
  search engines.
\item
  Twenty years ago, there was a smaller R community in which there
  was a common understanding of free software as a contract between developers and users to
  actively share in the progress of a given project (as in the spirit of the
  ``Debian Social Contract'', see \url{https://www.debian.org/social_contract}, especially point 2).
  Since then the R community has grown in size and complexity, the typical career
  paths of new members have changed, and R is increasingly perceived as one among many
  free-of-charge software applications. But in contrast to R, most of the other
  free-of-charge software is actually paid for by surrendering data, not by sharing
  a responsibility for the progress of the project.
\end{enumerate}

Both of these points have probably affected the expectations of users with regard to how
much effort to apply to learning about the software they have chosen to use (but
which they are not purchasing). This needs to be taken into account in further improvement
of the effectiveness of task views. Therefore, one goal is to improve search engine
optimization. Some first steps have been taken by adding more HTML meta information tags
(e.g, DublinCore, Highwire Press, Facebook, and Twitter) but more improvements are necessary.
Another goal is to better connect with those sub-communities for whom the task views are
relevant. This could be accomplished by including more representatives from these sub-communities
in broaders teams of task view maintainers, in order to restore a certain social contract
and sharing the responsibility for the task views. Moreover, listing the task views
in more online overviews/tutorials geared towards these different sub-communities
would help to spread the information (and also help with uptake by search engines).

We hope that the relaunch of the CRAN Task View Initiative will help us make progress
in these directions and we look forward to more activities and contributions in
the future!

\hypertarget{references}{%
\section*{References}\label{references}}
\addcontentsline{toc}{section}{References}

\hypertarget{refs}{}
\begin{CSLReferences}{1}{0}
\leavevmode\vadjust pre{\hypertarget{ref-ctvsuggest}{}}%
Dijk, Dylan. 2023. \emph{{CTVsuggest}: {CRAN} Task View Package Recommendations}. \url{https://dylandijk.github.io/CTVsuggest/}.

\leavevmode\vadjust pre{\hypertarget{ref-cran}{}}%
Hornik, Kurt, Uwe Ligges, and Achim Zeileis. 2022. {``Changes on {CRAN}.''} \emph{The R Journal} 14 (4): 356--57. \url{https://journal.R-project.org/news/RJ-2022-4-cran}.

\leavevmode\vadjust pre{\hypertarget{ref-bibentry}{}}%
Hornik, Kurt, Duncan Murdoch, and Achim Zeileis. 2012. {``Who Did What? {T}he Roles of {R} Package Authors and How to Refer to Them.''} \emph{The R Journal} 4 (1): 64--69. \url{https://doi.org/10.32614/RJ-2012-009}.

\leavevmode\vadjust pre{\hypertarget{ref-rforge}{}}%
Theußl, Stefan, and Achim Zeileis. 2009. {``Collaborative Software Development Using {R}-{F}orge.''} \emph{The R Journal} 1 (1): 9--14. \url{https://doi.org/10.32614/RJ-2009-007}.

\leavevmode\vadjust pre{\hypertarget{ref-knitr}{}}%
Xie, Yihui. 2015. \emph{Dynamic Documents with {R} and {knitr}}. 2nd ed. Boca Raton: Chapman \& Hall/CRC. \url{https://doi.org/10.1201/9781315382487}.

\leavevmode\vadjust pre{\hypertarget{ref-rmarkdown}{}}%
Xie, Yihui, J. J. Allaire, and Garrett Grolemund. 2018. \emph{{R} Markdown: The Definitive Guide}. Boca Raton: Chapman \& Hall/CRC. \url{https://doi.org/10.1201/9781138359444}.

\leavevmode\vadjust pre{\hypertarget{ref-ctv-intro}{}}%
Zeileis, Achim. 2005. {``{CRAN} Task Views.''} \emph{{R} News} 5 (1): 39--40. \url{https://CRAN.R-project.org/doc/Rnews/}.

\leavevmode\vadjust pre{\hypertarget{ref-ctv-pkg}{}}%
Zeileis, Achim, and Kurt Hornik. 2022. \emph{{ctv}: {CRAN} Task Views}. \url{https://CRAN.R-project.org/package=ctv}.

\end{CSLReferences}

\bibliography{ctv.bib}

\address{%
Achim Zeileis\\
Universität Innsbruck\\%
Department of Statistics\\
\url{https://www.zeileis.org/}\\%
\textit{ORCiD: \href{https://orcid.org/0000-0003-0918-3766}{0000-0003-0918-3766}}\\%
\href{mailto:Achim.Zeileis@R-project.org}{\nolinkurl{Achim.Zeileis@R-project.org}}%
}

\address{%
Roger Bivand\\
Norwegian School of Economics\\%
Department of Economics\\
\url{https://www.nhh.no/en/employees/faculty/roger-bivand/}\\%
\textit{ORCiD: \href{https://orcid.org/0000-0003-2392-6140}{0000-0003-2392-6140}}\\%
\href{mailto:Roger.Bivand@R-project.org}{\nolinkurl{Roger.Bivand@R-project.org}}%
}

\address{%
Dirk Eddelbuettel\\
University of Illinois at Urbana-Champaign\\%
Department of Statistics\\
\url{https://dirk.eddelbuettel.com}\\%
\textit{ORCiD: \href{https://orcid.org/000-0001-6419-907X}{000-0001-6419-907X}}\\%
\href{mailto:Dirk.Eddelbuettel@R-project.org}{\nolinkurl{Dirk.Eddelbuettel@R-project.org}}%
}

\address{%
Kurt Hornik\\
WU Wirtschaftsuniversität Wien\\%
Institute for Statistics and Mathematics\\
\url{https://statmath.wu.ac.at/~hornik/}\\%
\textit{ORCiD: \href{https://orcid.org/0000-0003-4198-9911}{0000-0003-4198-9911}}\\%
\href{mailto:Kurt.Hornik@R-project.org}{\nolinkurl{Kurt.Hornik@R-project.org}}%
}

\address{%
Nathalie Vialaneix\\
INRAE Toulouse\\%
Unité MIA-T\\
\url{https://www.nathalievialaneix.eu/}\\%
\textit{ORCiD: \href{https://orcid.org/0000-0003-1156-0639}{0000-0003-1156-0639}}\\%
\href{mailto:Nathalie.Vialaneix@inrae.fr}{\nolinkurl{Nathalie.Vialaneix@inrae.fr}}%
}

\end{article}

\end{document}